\documentclass[arguments]{aastex631}
\usepackage{amsmath}
\usepackage{xcolor}
\usepackage{footmisc}
\usepackage{rotating,longtable}
\usepackage{tablefootnote}
\usepackage{booktabs}
\usepackage{comment}

\usepackage{graphicx}
\usepackage{natbib} 
\usepackage{hyperref}
\usepackage{tablefootnote}
\usepackage{booktabs}

\usepackage{lipsum,mwe}

\received{}
\accepted{}
\begin{document}
\title{Low-mass stellar and substellar content of the young cluster Berkeley 59}
\author[0000-0002-0151-2361]{Neelam Panwar}
\affil{Aryabhatta Research Institute of Observational Sciences (ARIES),
Manora Peak, Nainital 263001, India}
\author{Rishi C.}
\affil{Aryabhatta Research Institute of Observational Sciences (ARIES),
Manora Peak, Nainital 263001, India}
\author[0000-0001-5731-3057]{Saurabh Sharma}
\affil{Aryabhatta Research Institute of Observational Sciences (ARIES),
Manora Peak, Nainital 263001, India}
\author[0000-0001-9312-3816]{Devendra K. Ojha} 
\affil{Tata Institute of Fundamental Research (TIFR), Homi Bhabha Road, Colaba, Mumbai - 400005, India}
\author[0000-0002-9431-6297]{Manash R. Samal}
\affil{Physical Research Laboratory, Navrangpura, Ahmedabad - 380009, India}
\author{H. P. Singh}
\affil{Department of Physics \& Astrophysics, University of Delhi, Delhi - 110007, India}
\author[0000-0002-6740-7425]{Ram Kesh Yadav}
\affil{National Astronomical Research Institute of Thailand (Public Organization), 260 Moo 4, T. Donkaew, A. Maerim, Chiangmai 50 180, Thailand}




\begin{abstract}
{
	We present a multi-wavelength analysis of the young star cluster Berkeley~59 based on the $Gaia$ data and deep infrared (IR) observations with the 3.58-m Telescopio Nazionale Galileo and $Spitzer$ space telescope. The mean proper motion of the cluster is found to be  $\mu$$_\alpha$cos$\delta$  $\sim$ -0.63 mas yr$^{-1}$ and $\mu$$_\delta$ $\sim$ -1.83 mas yr$^{-1}$ and the kinematic distance of the cluster, $\sim$ 1 kpc, is in agreement with previous photometric studies. Present data is the deepest available near-IR observations for the cluster so far and reached below 0.03 M$_\odot$. The mass function of the cluster region is calculated using the statistically cleaned color-magnitude diagram and is similar to the Salpeter value for the member stars above 0.4 M$_\odot$. In contrast, the slope becomes shallower ($\Gamma$ $\sim$ 0.01 $\pm$ 0.18) in the mass range 0.04 - 0.4 M$_\odot$, comparable to other nearby clusters. The spatial distribution of young brown dwarfs (BDs) and stellar candidates shows a non-homogeneous distribution. This suggests that the radiation feedback from massive stars may be a prominent factor contributing to the BD population in the cluster Be~59. We also estimated the star-to-BD ratio for the cluster, which is found to be $\sim$ 3.6. The Kolomogorov-Smirnov test shows that stellar and BD populations significantly differ, and stellar candidates are near the cluster center compared to the BDs, suggesting mass segregation in the cluster toward the substellar mass regime. 
}        
\end{abstract}



\section{Introduction} \label{intro}

Most of the stars in the Galaxy emerge out of the molecular clouds as clusters or groups \citep[e.g.,][]{lada03} with masses 
ranging from hundreds of solar masses extending down to the hydrogen burning limit. Despite being the focus of observational and theoretical research, several open issues remain in the field of star formation. One such issue is the dependence of star formation and early evolution physics on the mass. 
The low-mass stellar and substellar mass regimes are crucial to answering such questions and understanding the origin of the 
initial mass function (IMF) and the role of dynamical interaction, fragmentation, 
and accretion in determining the final mass of objects \citep{bonn06,bonnell08}. 

Substellar regime consists of objects that cannot sustain hydrogen fusion in their cores. Brown dwarfs (BDs) are the substellar mass objects having masses below 0.075  M$_\odot$  
extending down to about 0.015 M$_\odot$ that is well above the deuterium burning limit \citep[e.g.,][]{zapa00,lucas00,caballero18}, and overlap with the massive planets. 
Observations of star-forming regions (SFRs) also suggest the influence of stellar density on the number of very low-mass objects, with high-density regions producing more low-mass objects. 
 For example, an overabundance of low-mass objects is reported in the high-density region NGC~1333 \citep{scholz13,luhman16}, significantly different from the young cluster IC~348. Recent observations \citep[e.g.,][]{andersen08,muzic19,kubiak21} reveal a star-to-BDs ratio of $\sim$ 2 - 7 in different regions. 
\citet{andersen08} found that the ratios of star-to-BDs for different SFRs are consistent with a single underlying IMF. However, detailed analysis of the substellar content is primarily done in nearby SFRs that contain relatively loose groups of low-mass stars, except the Orion Nebula cluster (ONC). The IMF studies of the ONC yield somewhat inconsistent results, from a steeply declining IMF \citep{rio12} to 
a bimodal IMF with two peaks at 0.25 and 0.025 
 M$_\odot$ and dip at hydrogen burning limit \citep{drass16}. \citet{drass16} argue that the second IMF peak is possibly due to the BDs ejected from multiple systems or circumstellar discs at an early stage of star formation. However, \citet{mario20} do not observe a secondary peak in the BDs or planetary-mass regimes of the IMF, suggesting that the bimodal IMF of the ONC obtained using near-infrared (NIR) broadband observations may be due to the unaccounted background contamination. 
 
 The environmental influence on the production of substellar candidates can be probed by analyzing their census in different environments. However, reaching the substellar mass limit is challenging for young clusters as they are still embedded in the dust/ gas. Also, in some cases, crowding the sources in clusters may impose difficulty in resolving the individual sources in the cluster. Consequently, most studies of the stellar content in young star clusters extend to only above the hydrogen-burning mass limit. Observations of young clusters before they dynamically relax can indicate the initial conditions for forming their substellar content. 
 
 Berkeley 59 (Be~59) is a young cluster \citep[$\sim$ 2 Myr;][]{pan08} located at the centre of the Cepheus OB4 stellar association. It is a nearby cluster (distance $\sim$ 1 kpc) which is surrounded
by the H{\sc ii} region Sh2-171 \citep{yan92}. The reddening $E$($B$ - $V$) toward the cluster direction 
ranges from 1.4 to 1.8 mag \citep{maj08,pan08}.
It contains several massive stars (spectral type from O7 to B5) \citep{kun08,maj08,skif14,mai16,gahm22}
and associated natal molecular clouds, making it one of the potential nearby massive clusters \citep{panwar18}. 
\citet{yan92} observed two dense molecular clumps (`C1' and `C2') at the western side of the Be~59 using
the J = 1 - 0 lines of $^{12}$CO and $^{13}$CO emission. They suggested that the dense gas is in contact with
the H{\sc ii} region Sh2-171, and the ionization front (IF) is driving shocks into the clumps, and a new generation of stars may be triggered to form from the compressed gas layer. \citet{ros13} studied the region using Wide-field Infrared Survey Explorer telescope  
data and identified a small group of embedded stars (RA: 00h 00m 46s, DEC: +67$^\circ$32$^\prime$59$\arcsec$, J2000.0) which resides in a cloud within the clump `C2'. 

The cluster has been investigated by several authors at
optical bands \citep{maj08,pan08,lat11,esw12}; most of these studies were limited to the high to intermediate-mass
stars only. \citet{pan08} identified the intermediate mass population of young stellar objects (YSOs) in the region 
using NIR 2 micron all sky survey (2MASS) and slitless grism H$\alpha$ data. However, their optical photometry could detect stars up to V $\sim$ 18.5 mag corresponding to $\sim$ 1.5 M$_\odot$ at the adopted distance and extinction of the cluster. \citet{panwar18} characterized the low-mass young stellar population of the cluster by using deep optical photometric data. They studied the IMF for the cluster region to understand the nature of the star formation process and the properties of stellar systems. X-ray, NIR, and mid-infrared (MIR) data analysis revealed more than 600 YSO candidates in the region \citep{koe12,ros13,get17}. 
Therefore, Be~59 is a nearby cluster containing O/B stars and YSOs and thus provides an ideal environment to look for very low-mass stars and the substellar candidates and to examine the potential differences in the BD formation efficiency. 

We present deep NIR observations of the central region ($\sim$ 2.5 $pc$ $\times$ 2.5 $pc$)
of the young cluster Be~59 obtained with the 3.58-m Telescopio Nazionale Galileo (TNG). 
Previous deep optical observations of the same central portion of the cluster could reach up to $\sim$ 0.1 M$_\odot$ \citep{panwar18}, 
whereas the present data extend further into the substellar mass regime. These new observations are the deepest to date for this
region and allow us to assess the stellar contents of the
cluster down to the hydrogen burning limit (0.075 M$_\odot$) and provide a better analysis of its properties and mass function, understanding of the dynamical status, and insight into its formation. 
In Section 2, we present the NIR observations, the data reduction, and the completeness of the NIR dataset. In Section 3, the identification 
of the young stellar population, kinematic members, and distance of the cluster are discussed. Section 4 discusses the IMF and clues on the formation of a substellar population. 
Finally, the results of the present work are summarized in Section 5.

\section{Observations and Data Reduction}
\begin{figure}
\centering
\includegraphics[scale=0.55, trim = 50 0 0 0, clip]{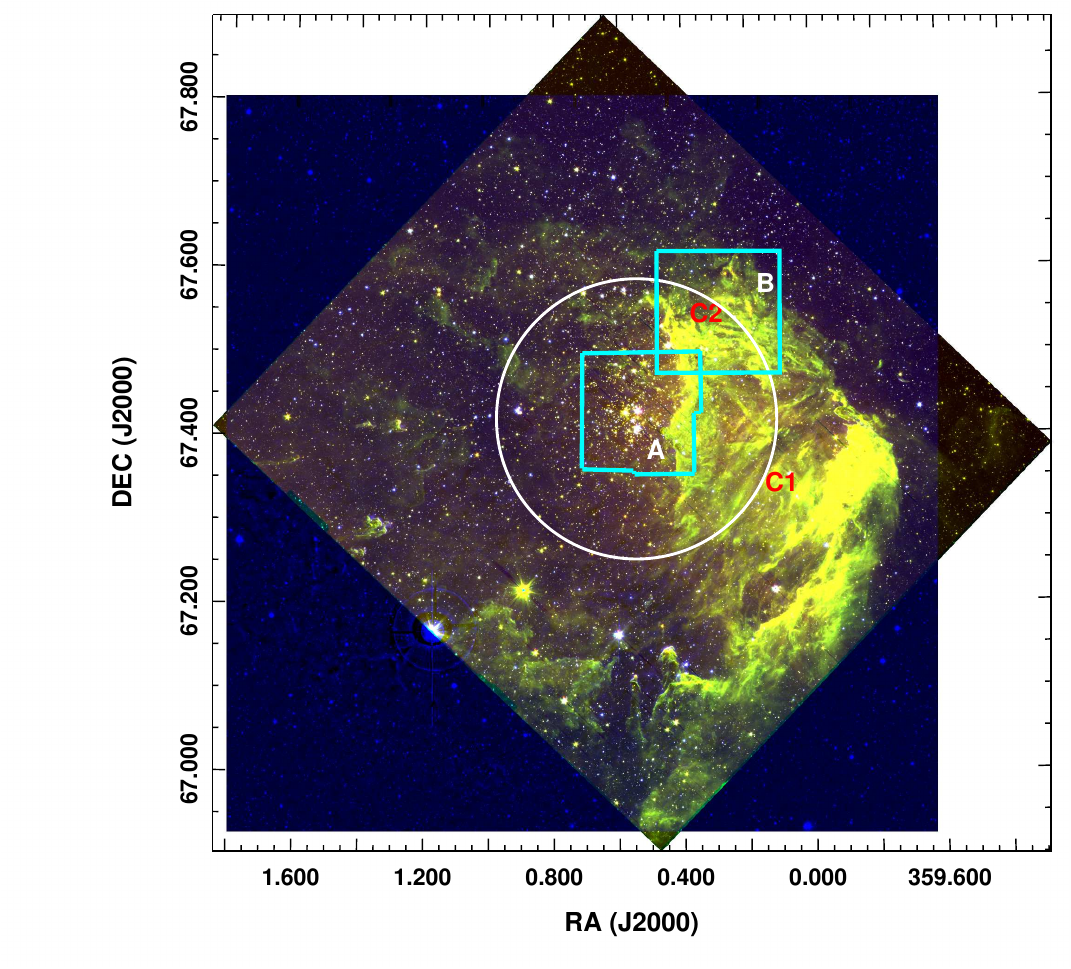}
\caption{Three-color image (red: 4.5 $\micron$, green: 3.6 $\micron$, blue: CFHT MegaCam $r$-band) of the 
Be~59 region. The two boxes (marked as `A' and `B')  
represent the regions covered with the NICS observations.  The white circle represents the extent of the cluster \citep[$\sim$10$^\prime$,][]{pan08}. 
`C1' and `C2' are the locations of the two dense clumps identified by \citet{yan92}. 
}
\label{fig1}
\end{figure}

 Fig. \ref{fig1} shows the color-composite image (red: 4.5 $\micron$, green: 3.6 $\micron$ from Spitzer-IRAC , blue: $r$-band 
Canada–France–Hawaii Telescope (CFHT) MegaCam image) of the Be~59 region.  
The two boxes of $\sim$ 8$^\prime$.6 $\times$ 
8$^\prime$.6 each represents the area covered by our NIR observations. The young cluster Be~59 is at the center of the image. The white circle represents the extent of the cluster \citep[$\sim$10$^\prime$,][]{pan08}. `C1' and `C2' represent the locations of the peaks of the molecular clumps \citep[cf.][]{yan92}. 
The diffuse optical emission is mainly seen at the cluster's center, indicating
low extinction at the central portion of the cluster compared to the outskirts, especially 
towards the west, where emission in 3.6 $\micron$ and 4.5 $\micron$ is present. In this work, we used the deep NIR $JHK$ observations of two sub-regions of the cluster to examine the physical properties of the cluster Be~59, identify the low-mass stellar and substellar population, and understand the origin of these young stellar/ substellar sources.  
\subsection{Near-infrared Observations}
The deep {\it JHK} observations of the two sub-regions of the cluster Be~59 (see Fig. \ref{fig1}), region `A' (RA: 00h 02m 14.0s, DEC:+67$^\circ$25$^\prime$08$\arcsec$.0, J2000.0) 
and region `B' (RA: 00h 01m 08.0s, DEC:+67$^\circ$33$^\prime$03${\arcsec}$.0, J2000.0), 
and a control field having the same area (centered at RA: 00h 09m 48.0s, DEC: +68$^\circ$07$^\prime$58$\arcsec$.0 , J2000.0) were obtained by 
using the Near Infrared Camera and Spectrograph (NICS) mounted on the 3.58-m Italian Telescope TNG at the Observatorio de los Muchachos in La Palma (Canary Islands, Spain) during October/ November, 2010. NICS is a 1024 x 1024 HgCdTe HAWAII near-infrared array detector (with a plate scale of 0$^{\prime\prime}$.25/pixel) and 
covers a field of view $\sim$ 4$^\prime$.2 $\times$ 4$^{\prime}$.2 \citep{baffa01}. To cover the regions `A' and `B', we observed a raster of 
4 $\times$ 4 pointings, and at each pointing, we obtained nine dithered exposures of 45 seconds each. During the observations, 
the average seeing was $\sim$ $1^{\prime\prime}$.

The pre-processing of the NIR images required additional steps, and due to the bright background in the NIR, the sky contribution from the target images had to be removed. 
The NICS images need to be corrected for cross-talk (i.e., a signal detected 
in one quadrant, producing negative ghost images in the other three quadrants) and for the distortion of the NICS optics. 
These corrections were performed using the Speedy Near-IR data Automatic reduction pipeline (SNAP) 
pipeline developed by F. Mannucci, available at TNG to reduce NICS data. 
The source detection and NIR photometric measurements of the detected sources in the processed frames were performed by using $DAOPHOT-II$ 
software package \citep{ste87}. The point spread function (PSF) was obtained for each 
frame by using several uncontaminated stars, and the PSF photometry of all the sources was obtained using 
the ALLSTAR task in $IRAF$.

The instrumental magnitudes were converted to the standard 2MASS system using the 2MASS $JHK$ data \citep{cut03}. 
The photometric accuracies depend on the brightness of the stars. We finally consider only those sources 
with uncertainty $<$0.1 mag in $J$, $H$, and $K$ bands. Our data (detection limit, $K$ $\sim$ 18 mag) are $\sim$ 2.5 mag deeper than the available NIR 2MASS data. 

\subsubsection{Data Completeness}\label{compl}
To derive cluster properties, it is necessary to take into account the 
incompleteness in the observed data that may occur for various reasons, e.g., crowding of the stars, diffuse emission, variable extinction, etc. We used the $ADDSTAR$ routine of $DAOPHOT-II$ to artificially add the randomly positioned 
stars and determine the completeness factor (CF). 
The procedure has been outlined in detail in our earlier works \citep[e.g.,][]{pan01,cha11}. 
We randomly added the artificial stars to the $J$, $H$, and $K$ band images so that they 
have similar geometrical locations. The luminosity distribution of artificial stars is chosen 
to insert more stars towards the fainter magnitude bins. The frames were 
reduced using the same procedure used for the original frame. The ratio of the number of 
stars recovered to that added in each magnitude interval gives the CF as a function of magnitude. 
\begin{figure}
\centering
\includegraphics[scale = 0.5, trim = 0 0 0 0, clip]{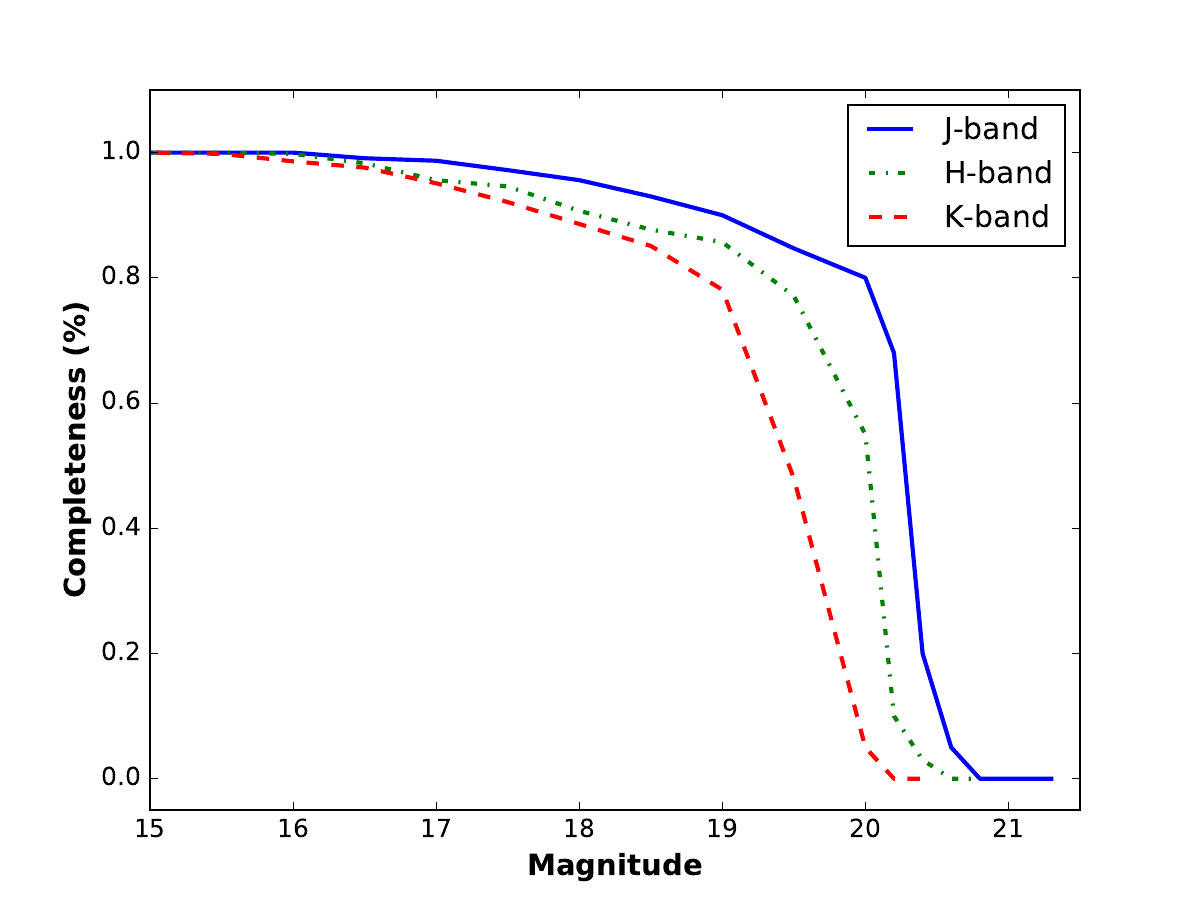}
\caption{Completeness of the NICS photometry in $J$, $H$ and $K$ bands calculated with an artificial star experiment.}
\label{fig2}
\end{figure}
These values were used to correct the data for incompleteness. The incompleteness of the data 
increases with an increasing magnitude, as expected. 

The PSF photometry and completeness analyses were performed similarly on the control field images. The results of our 
completeness analyses in the cluster region are shown in Fig. \ref{fig2}.
\subsection{Previously Known YSO candidates}
During their early phase, stars possess circumstellar disks, showing excess emission in the IR 
wavelengths. \citet{get17} carried out the {\it Star Formation in Nearby Clouds (SFiNCs)} project 
to provide detailed studies of the young stellar populations and star cluster formation in nearby SFRs, including Be~59. They identified $\sim$ 700 young stars using 2MASS ($JHK$), 
$Spitzer$-IRAC (3.6 $\micron$, 4.5 $\micron$) and X-ray data 
(from $Chandra$ space telescope) and classified them as sources with disks (Class {\sc i/ii}), without disks (Class {\sc iii}) and probable members. 
Our target regions are well within their study area, so we have used their YSO catalog in the present work. 

\subsection{Complimentary archival datasets}
\subsubsection{Gaia Data Release 3}
We use the Gaia data release 3 (Gaia DR3) \citep{gaia22}, which contains positions, proper motions, and parallaxes values. We note that Gaia DR3 data are very useful for accurate distance estimation to the cluster. Still, these are limited to only relatively bright low-mass stars ($G$ $\sim$ 20 mag) as fainter stars have large uncertainties in measurements.
\subsubsection{Optical $I$-band data from previous work}
Combining the NIR data with the optical data is crucial for accurately identifying ultracool late-L and T-type dwarfs and distinguishing them from photometric contaminants \citep{ramirez12}. As the same region is also observed with the TNG in the optical $VI$ bands, we have used the $I$-band measurements for the stars having deep NIR observations from \citet{panwar18}. Our $I$-band data are available for the stars with $I$ $\sim$ 23 mag.

\section{Results}
\begin{figure}
\centering
\includegraphics[scale = 0.55, trim = 0 0 0 0, clip]{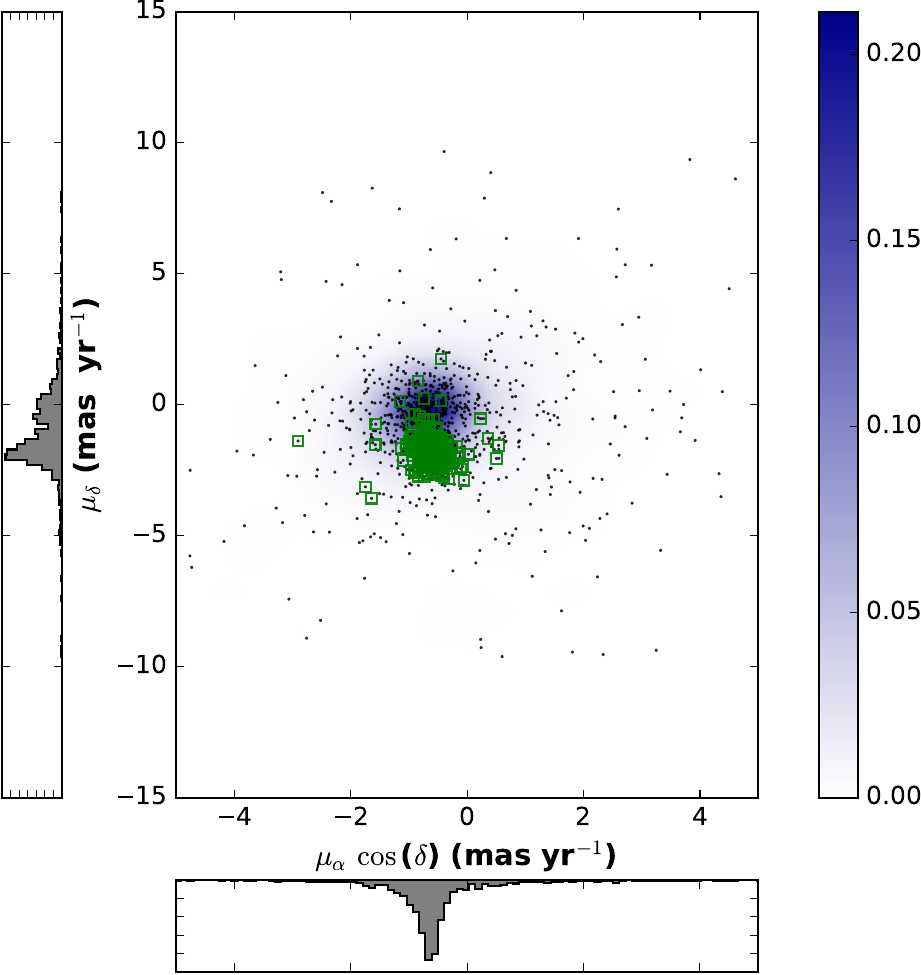}
	\caption{Vector-point diagram for the stars in the cluster region (radius $\le$ 10$^\prime$) is shown with the black dots. The distribution of the stars in a control field of a similar area, a two-dimensional kernel density estimation eKDE, is shown with the colorbar. Green squares represent the YSOs identified by \citet{get17}.}
\label{fig3}
\end{figure}
\subsection{Kinematic members with the Gaia DR3 data: Proper motion, membership and distance of the stars in the cluster region}
\begin{figure}
\centering
\includegraphics[scale = 0.4, trim = 0 0 0 0, clip]{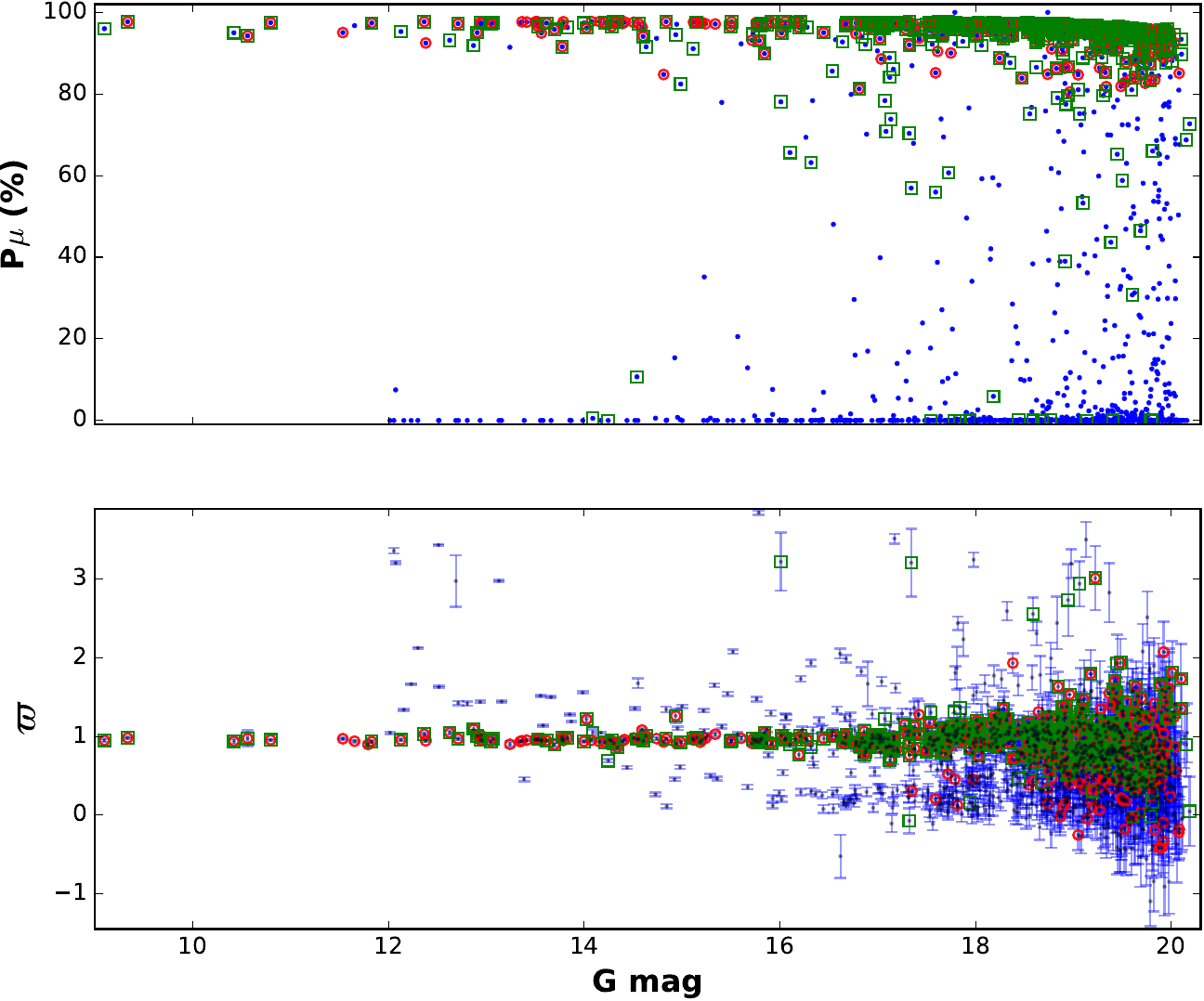}
\caption{The membership probability (top panel) and parallax values (bottom panel) for the stars in the cluster region as a function of the $G$-magnitude (blue dots). Red circles show the members with $P$$_\mu$ $>$ 80\%, and green squares represent the YSO candidates from \citet{get17}.  }
\label{fig4}
\end{figure}

\begin{figure}
\centering
\includegraphics[scale = 0.4, trim = 0 0 0 0, clip]{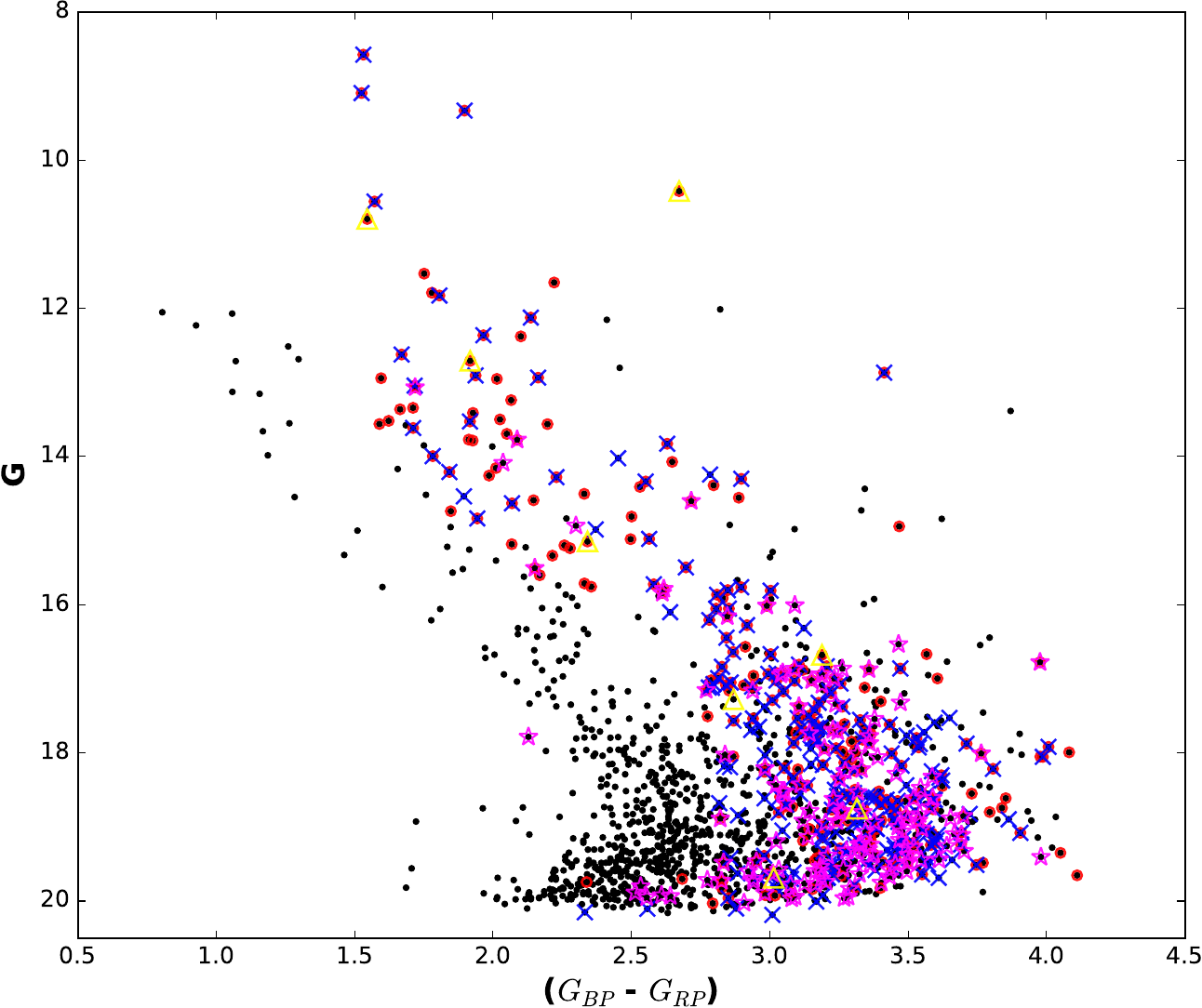}
\caption{Gaia DR3 CMD for the stars in the cluster region. The red circles are the member stars with high membership probability ($>$ 80\%) within the proper motion ellipse shown in VPD and parallax values within 2$\sigma$ of the 
median. Known YSO candidates from \citet{get17} are shown with the blue cross (Class {\sc iii}), magenta star (Class {\sc ii}), and yellow triangles
(probable members).}
\label{fig5}
\end{figure}

For the kinematic analysis of the cluster, we used the Gaia DR3 data of the cluster region (radius $\sim$ 10 arcmin). We used the positions (RA and Dec), proper motions ($\mu$$_\alpha$cos($\delta$) and $\mu$$_\delta$), photometric magnitudes ($G$, $G_{BP}$, $G_{RP}$), parallax ($\varpi$) measurements and their associated uncertainties of the stars having renormalized unit weight error ($RUWE$) $>$ 1.4. 
The proper motions (PMs) of the sources within the cluster region and having magnitude uncertainty $<$ 0.1 mag in the $G$-band are 
plotted in the vector point diagram (VPD) shown in Fig. \ref{fig3}. In Fig. \ref{fig3}, we have also shown the histograms for the $\mu$$_\alpha$cos($\delta$) and $\mu$$_\delta$. The histogram distributions of proper motion values indicate two distinct groupings of sources, one around $\mu$$_\alpha$cos($\delta$),$\mu$$_\delta$ $\sim$ (0.63, -1.0) mas yr$^{-1}$ and another at $\mu$$_\alpha$cos($\delta$),$\mu$$_\delta$ $\sim$ (-0.63, -1.83) mas yr$^{-1}$. Therefore, it is essential to confirm which of these two groups belongs to the kinematic members of the cluster. For this, we examine the distribution of the PMs of the sources within the control field region of a similar area (shown with the blue color in Fig. \ref{fig3}). 
By comparing the PM distributions of the sources within the cluster region and control field, it is clear that most of the stars in the control field occupy the former group, and there are only negligible field sources in the latter group. 
Hence, we consider the $\mu$$_\alpha$cos($\delta$),$\mu$$_\delta$ $\sim$ (-0.63, -1.83) mas yr$^{-1}$ for the  cluster. 

We also estimated the membership probability for the sources within the cluster region (radius $\le$ 10 arcmin). 
By assuming a distance of $\sim$ 1 kpc and a radial velocity dispersion of 1 km s$^{-1}$ for open clusters 
\citep{girard89}, a dispersion ($\sigma$$_c$) of $\sim$ 0.21 mas yr$^{-1}$ is expected. 
For the remaining field stars, we have calculated $\mu$$_{xf}$= -0.42 mas yr$^{-1}$, $\sigma$$_{xf}$ = 1.13 mas yr$^{-1}$ 
and $\mu$$_{yf}$= -0.54 mas yr$^{-1}$, $\sigma$$_{yf}$ = 2.04 mas yr$^{-1}$, the mean and standard deviation of their PM values in the RA and Dec, respectively. These values are further used to construct the frequency distributions of cluster stars 
($\phi^\nu_c$) and field stars ($\phi^\nu_f$) by using the equations given in \citet{yadav13} and then the value of membership probability (ratio of the distribution of cluster stars with all stars) for the $i^{th}$ star is estimated as:

\begin{equation}
{P_\mu(i)} = {{n_c\times\phi^\nu_c(i)}\over{n_c\times\phi^\nu_c(i)+n_f\times\phi^\nu_f(i)}}
\end{equation}
 where n$_c$(=0.38) and n$_f$(=0.62) are the normalized numbers of stars for the cluster and field regions (n$_c$ + n$_f$=1). 
 The estimated membership probability of the Gaia sources located within the radius of the Be~59 is 
 plotted as a function of $G$-magnitude in Fig. \ref{fig4} (top panel). As seen in this plot, 
 a high membership probability ($P_\mu$ $>$ 80\% shown with red circles) extends down to $G$ $\sim$ 20 mag whereas, at fainter magnitudes, the probability 
 gradually decreases. Except for a few outliers, most of the stars with high membership probability ($P_\mu$ $>$ 80\% 
and $G$ $<$ 20 mag) follows a tight distribution. In summary, from the above analysis, we calculate the membership 
probability of $\sim$ 1520 stars in the Be~59 cluster region, and $\sim$ 460 stars were assigned as cluster members based on their high membership probability.

In Fig. \ref{fig4} (bottom panel), we have also plotted the parallax ($\varpi$) of the same stars as a function of $G$-magnitude. 
 The respective uncertainties in the parallax values are shown with the error bars. The red circles represent sources having P$_\mu$ $>$ 80\%. 
To validate the distance of the cluster, we used the median distances of the 
high-membership probability ($P_\mu$ $>$ 80\%) cluster members with good parallax values (parallax error $<$ 0.1 mas yr$^{-1}$), provided by \citet{bailler21}. 
The distance for the Be~59 cluster using the Gaia data is estimated as $\sim$ 1.00 $\pm$ 0.06 kpc, which agrees with that reported by \citet{pan08}.

From the above analysis, it is clear that using the Gaia DR3 data, our sample of kinematic members of the cluster is restricted to only relatively bright members ($G$ $<$ 20 mag). Fig. \ref{fig5} shows the color-magnitude diagram (CMD) for the stars in the cluster region, constructed using Gaia DR3 data (black dots). The red circles represent high membership ($P_\mu$ $>$ 80\%) sources that are 
located within the ellipse shown in Fig. \ref{fig3} and have their parallax values within 2$\sigma$ of the median parallax. 
The probable member sequence stands out to the right of the bulk of the ﬁeld star population, except toward the fainter end, where the separation between field sources and member stars is unclear. Like ours, we find that the Gaia CMD for the Be~59, constructed by  \citet{perren23} from the unified cluster catalog, also shows that most of the stars have lower membership probability towards the fainter-end.  
We also looked for the young stellar candidates of \citet{get17} in Gaia DR3 data using a matching radius of $\sim$ 1$^{\prime\prime}$ and found $\sim$ 370 young stellar candidates within the cluster radius. These sources are shown with the crosses (Class~{\sc iii}), star symbols (Class~{\sc ii}), and yellow triangles (probable members) in the Gaia CMD. The distribution of these YSO candidates is similar to the member stars selected using kinematic information. However, we notice that towards the fainter end, there is no clear separation between cluster and field population, and therefore, the cluster population may be contaminated with the field stars. Thus, in the ensuing section, we adopt a photometric approach to minimize the contribution of contaminants in the cluster population.

\subsection{Photometric members of the cluster}
\begin{figure}
\centering
\includegraphics[scale = 0.5, trim = 0 0 0 0, clip]{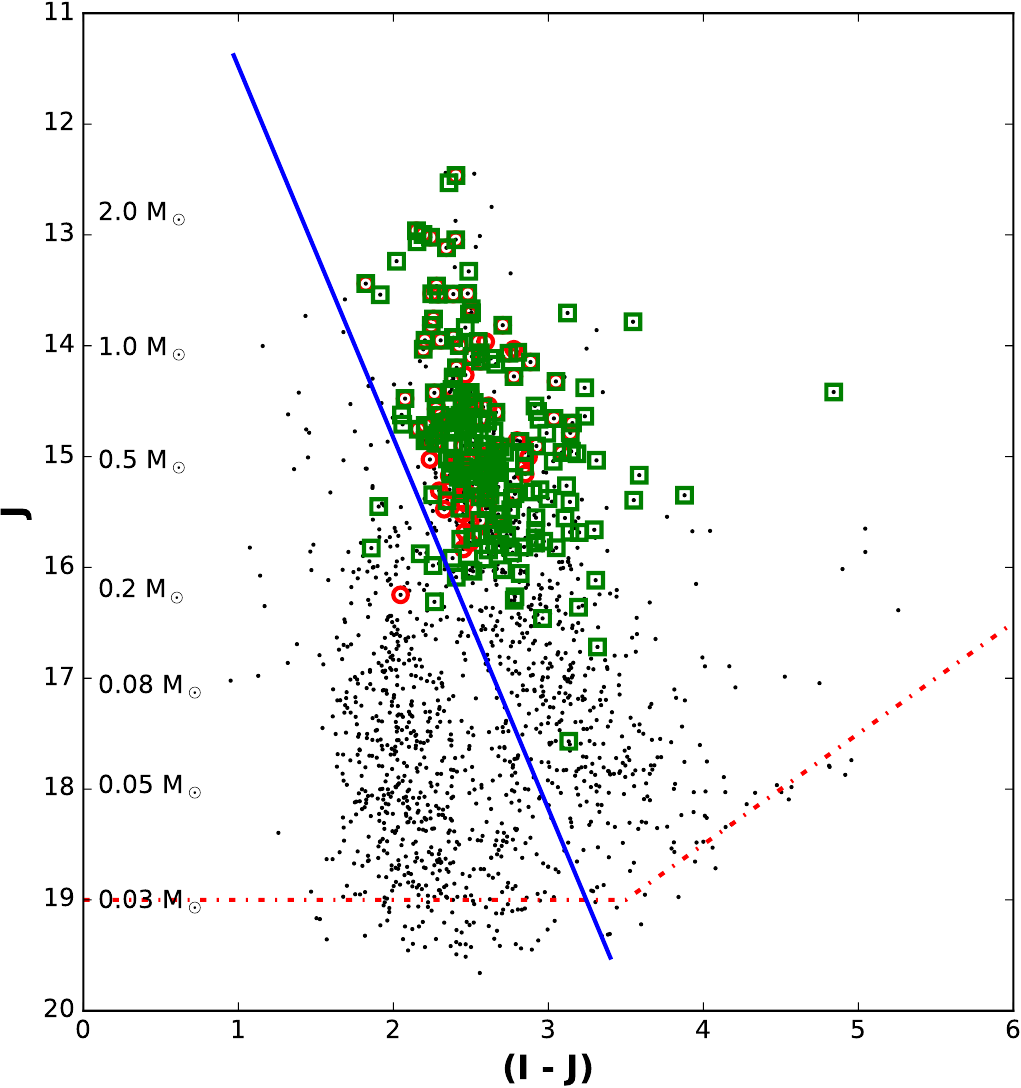}
\caption{Optical-NIR CMD for the stars in the region `A'. The red circles are the member stars selected based on the Gaia DR3 data. Known YSO candidates from \citet{get17} are shown with green squares (probable members). The red dashed line represents the completeness of the data.}
\label{fig6}
\end{figure} 
The members selected based on the Gaia DR3 kinematic and photometric data trace the cluster sequence up to J$\sim$ 16 mag. As the main focus of the present work is to identify cool objects that prominently have red colors,  we have utilized various photometric color-magnitude and color-color diagrams to obtain these objects.
\subsubsection{The optical/ NIR color-magnitude diagram}
We constructed the optical/ NIR CMD to identify the cluster's faint cool members. We merged our 2MASS/ NICS catalog with the $I$-band DOLORES photometry from \citet{panwar18} and found about 1490 sources having $I$ and $J$ band measurements with uncertainties in magnitudes $<$ 0.1 mag. 
Fig. \ref{fig6} shows the $J$/($I$ - $J$) CMD for the stars in the region `A'. We also search for the YSO candidates from \citet{get17} and Gaia members in our DOLORES/ NICS catalog. We found that $\sim$ 200 YSO candidates \citep{get17} have counterparts in our optical/ NIR catalog. In Fig. \ref{fig6}, red circles and green squares represent the Gaia members and YSOS from \citet{get17}, respectively. The CMD seems to contain two stellar populations, one toward the left of the blue line and the other right. Most YSO candidates and Gaia members have ($I$ - $J$) colors $>$ 2 mag. This manifests that the population towards the right of the blue line belongs to the cluster, whereas the former relates to the field population. We traced a
blue envelope along the known cluster members sequence in the CMD to select new candidate members, plotted as a thick blue line in Fig. \ref{fig6}. This line separates field sources
from the cluster member candidates and roughly corresponds to a linear fit to the cluster sequence in the $J$ vs. ($I$ - $J$) CMD. Adopting this criteria, out of 232 member candidates, 226 were retained. 
Therefore, by comparing the number of field stars to that of stars located right of the thick blue line, the level of contamination by field stars in our YSO sample is less than 3\%. 

The CMD sample obtained through this criteria seems to be the deepest. Still, it may not be the most complete sample and contains an enhanced proportion of contaminants from reddened background sources and embedded stars. Therefore, additional selection criteria discussed in the following subsections are utilized to select young and low-mass members.

\subsection{Additional disk bearing sources: Infrared excess emission}
\subsubsection{NIR excess emission}
Young disk-bearing stars emit excess emission in IR wavelengths that can be visible by examining their location in the NIR color-color diagram.  
Adopting the approach discussed in \citet{panwar20}, using the NIR ($J$ - $H$)/($H$ - $K$) color-color diagram, we identify the IR excess stars in the regions. 
The IR excess stars identified in the present work using TNG and 2MASS data are shown with the open triangle symbols in Fig. \ref{nircmd} (left panel).

\subsubsection{Young stellar candidates from 2MASS/NICS and Spitzer-IRAC data}
Since the current NIR data are substantially deeper than the 2MASS survey, we utilize the near-infrared $JHK$ measurements along with the $Spitzer$ 3.6 and 4.5 $\micron$ measurements to obtain additional young sources with excess emission in the cluster.
To identify the IR excess sources using $H$, $K$, 3.6 $\micron$ and 4.5 $\micron$ photometry, we have used the approach discussed in \citet{pan17}. To estimate the intrinsic NIR magnitudes and colors of the sources in the region, we created the extinction map of the region as discussed in \citet{pan17}. Further, we used the intrinsic color-cuts mentioned in \citet{gut09} to select the Class~{\sc ii} and Class~{\sc 0/i} sources. Using this approach, we identified about 679 Class~{\sc ii} and 57 Class~{\sc 0/i} sources. These sources are shown with star symbols in Fig. \ref{nircmd}.
The additional young sources identified in the present work are given in Table 1.
Only a portion of Table 1 is given here, and the complete table is available in
the electronic version.
\begin{table*}
\caption{$JHK$ magnitudes of the young sources }
\begin{tabular}{cccccc} \hline
	S. No. &RA (J2000)& Dec (J2000)& $J$ $\pm$ e$J$ & $H$ $\pm$ e$H$ & $K$ $\pm $e$K$\\
\hline
1&0.683588&67.25895 &14.99 $\pm$ 0.05 & 13.49 $\pm$ 0.04&12.29 $\pm$ 0.03 \\
2&0.522299&67.29337 &14.61 $\pm$ 0.04 & 12.83 $\pm$ 0.04&11.78 $\pm$ 0.02 \\
3&0.492325&67.31194 &16.03 $\pm$ 0.09 & 13.99 $\pm$ 0.05&12.73 $\pm$ 0.03 \\
\hline    
\end{tabular}
\end{table*}

\subsubsection{Color-magnitude diagram: Identification of young cluster members down to brown dwarf limit}
The CMDs are ideal for confirming the cluster members and determining their mass and evolutionary stages. 
However, these may be contaminated by the field population that is present along the line of sight 
\citep[e.g.,][]{sung10,da10,cha11,sung13,pan13a}. The cluster population can be distinguished from the general Galactic population along the line of sight through a comparison of the CMDs of the cluster region and the control field \citep[e.g.,][]{bran08,shar08,gen11,hay15}. 

\begin{figure*}
\centering
\includegraphics[scale = 0.73, trim = 0 0 0 0, clip]{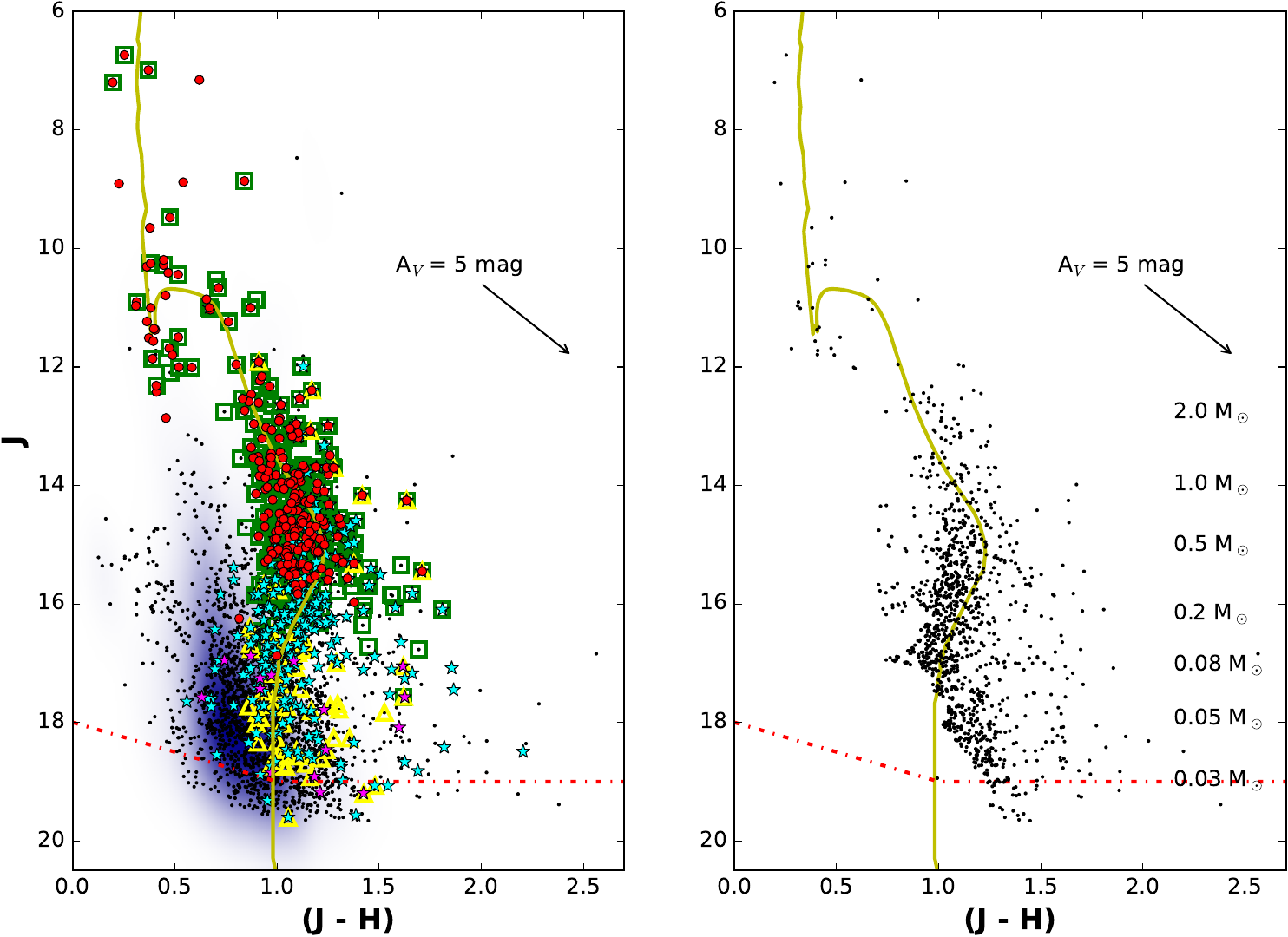}
\caption{Left panel:  $J$ vs. ($J$ - $H$) CMD for the sources in the region `A'. Green squares represent the YSO candidates identified by \citet{get17}. Yellow triangles are the NIR excess sources selected using the NICS observations (see sec. 3.3.1). Star symbols are the Class {\sc i/ii} sources selected based on the NICS/IRAC 3.6, 4.5 $\micron$ data. The distribution of the stars in the control field of the same area is shown in blue. Right panel: Statistically cleaned $J$ vs. ($J$ - $H$) CMD for the region `A' sources. The yellow curve is the 2 Myr isochrone from PARSEC \citep{bressan12,pastor20} and  \citet{baraffe15} corrected for the adopted distance and reddening. The red dashed line shows the completeness limit. }
\label{nircmd}
\end{figure*}

The NIR CMD can be crucial in identifying a population of YSO candidates, as these are easily distinguishable 
from the MS stars. In Fig \ref{nircmd} (left panel), we show the  $J$ vs. ($J$ - $H$) CMDs for the sources in the cluster region `A' with the black dots. The thick curve overplotted on the CMD represents 2 Myr isochrone from \citet{pastor20} and \citet{baraffe15} for $>$1 M$_\odot$ and $\le$ 1M$_\odot$, respectively. The isochrone is scaled for a visual extinction $A_V$ $\sim$ 4 mag and distance of $\sim$ 1 kpc towards the cluster region, derived in the previous work \citep{panwar18}. 
Fig. \ref{nircmd} (left panel) shows that the present NIR data reach well below the hydrogen burning limit and reveal sources less than 0.03 M$_\odot$. 
The distribution of sources in the region `A' on the CMD clearly demonstrates two separate sequences up to $J$ $\sim$ 16 mag. However, towards the fainter end, these two become indistinguishable. 
Therefore, to identify a sequence representing the cluster members, we also overplotted the kinematic members selected using Gaia data (red-filled circles), previously identified YSO candidates by \citet{get17} (green squares), and the young IR excess sources identified in the present work (star symbols and triangles) on the CMD (see Fig. \ref{nircmd}). In doing so, we found that most YSO candidates and cluster members have redder colors. 
Further, we also examined the distribution of the control field stars' on the CMD (shown with blue color in Fig. \ref{nircmd}). 
As can be seen, the cluster population stands out in the CMD  up to $J$ $\sim$ 16 mag (corresponding to $\sim$ 0.2 M$_\odot$). However, towards the low-mass end, there is likely contamination by the field stars. 
Therefore, to minimize the contaminants towards the fainter end, we considered young sources selected based on the NIR and/or IRAC observations as substellar cluster members.

\begin{figure*}
\centering
\includegraphics[scale = 0.64, trim = 0 0 0 0, clip]{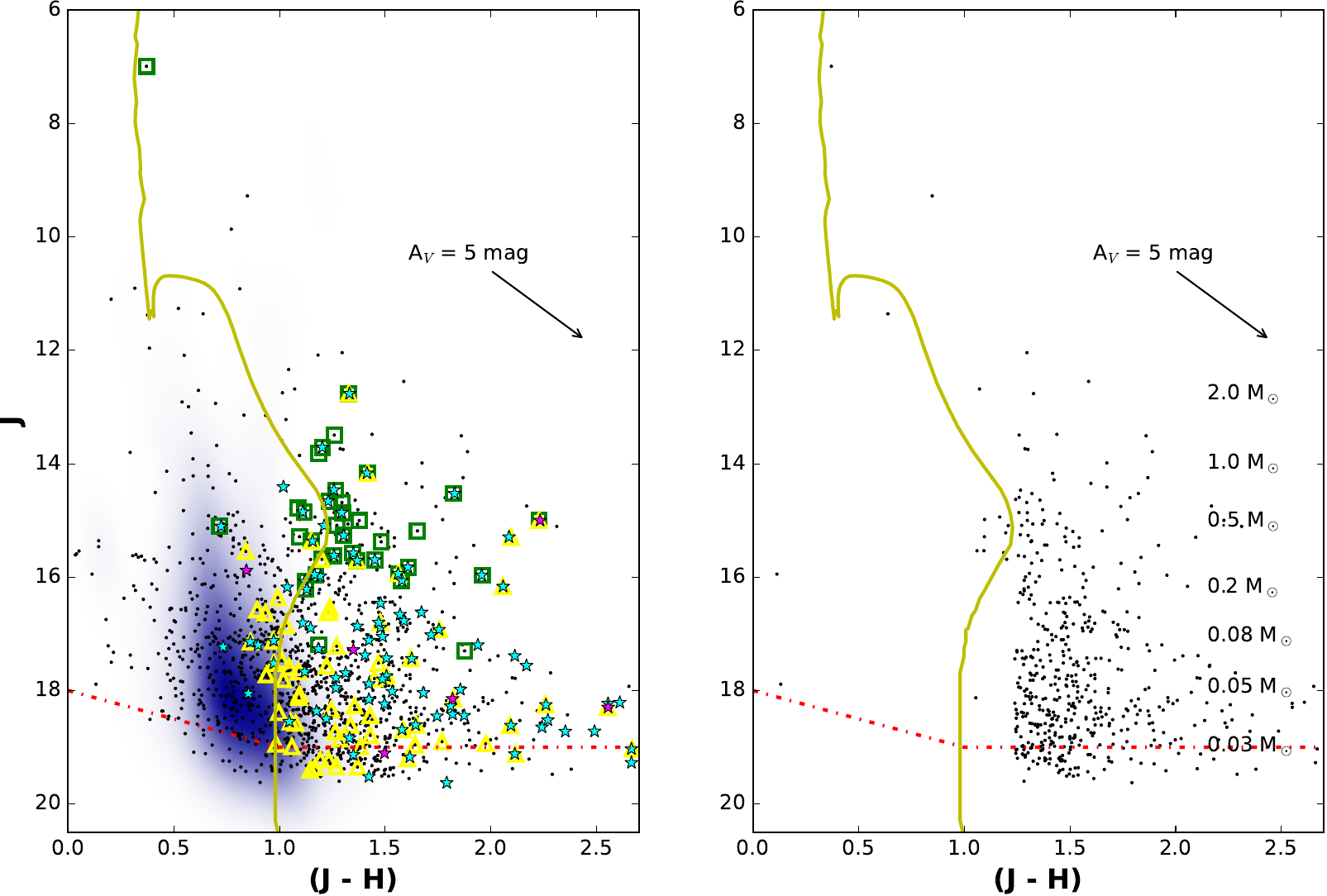}
\caption{Left panel:  $J$ vs. ($J$ - $H$) CMD for the sources in the region `B'.  Right Panel: Statistically cleaned $J$ vs. ($J$ - $H$) CMD for the sources in the region `B'.  The yellow curve is the 2 Myr isochrones from PARSEC \citep{bressan12,pastor20} and  \citet{baraffe15} scaled for the adopted distance and reddening. Other symbols are the same as in Fig. \ref{nircmd}.}
\label{cleancmd2}
\end{figure*}

We have also plotted the NIR CMD for the sources in the region `B' (see Fig. \ref{cleancmd2}). In Fig. \ref{cleancmd2} (left panel), the young stellar and substellar candidates identified from the previous and current works are also shown. The symbols in the Figure are the same as in Fig. \ref{nircmd}.
To estimate the masses of young low-mass sources, we utilized the NIR CMD. We traced back each source to the 2 Myr isochrone along the reddening vector and estimated the corresponding mass. While conducting our analysis, we incorporated random errors in mass estimation resulting from photometric measurements by employing Monte Carlo simulation. We created 1000 magnitudes and colors for each source, assuming a normal distribution with a standard deviation equivalent to the respective photometric uncertainty. This allowed us to estimate the average mass and its uncertainty for each source. The CMDs for the sources in region `A' and region `B' reveal that both regions contain young low-mass stellar and substellar objects. The order of errors in the mass estimation is $\sim$ 0.01 M$_\odot$ for the low-mass stellar and substellar sources ($<$ 1 M$_\odot$).


\begin{figure}
\centering
\includegraphics[scale = 0.48, trim = 0 0 0 0, clip]{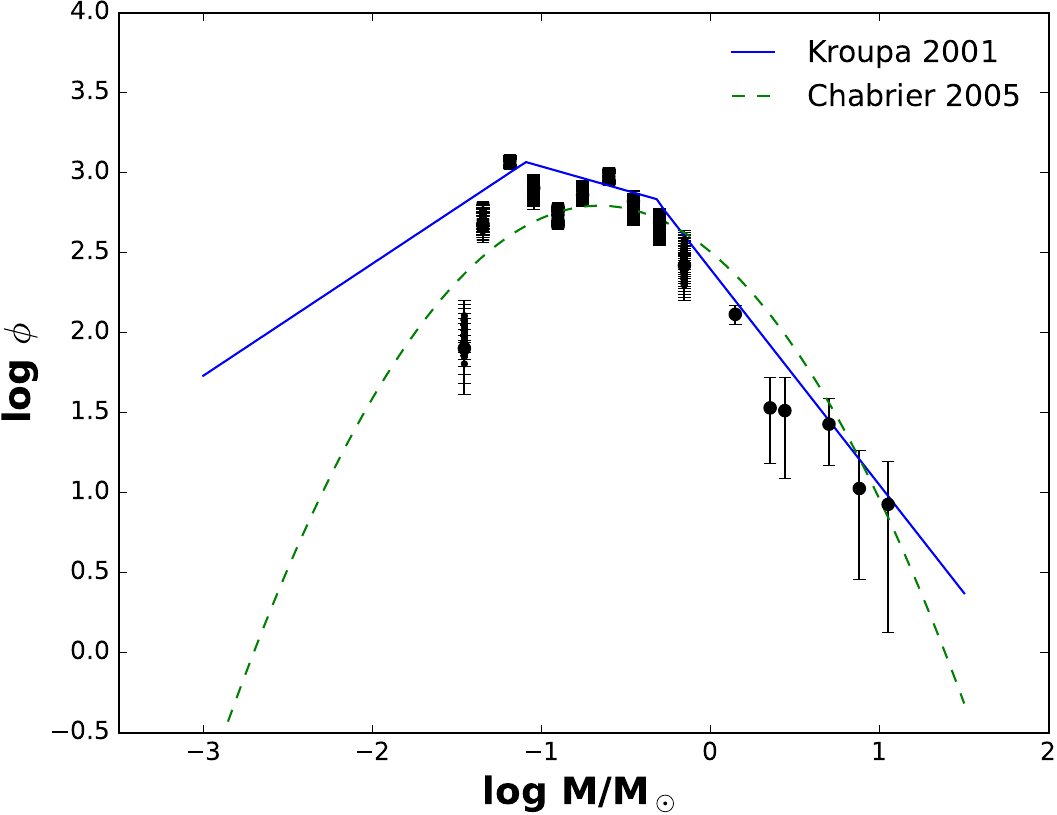}
\caption{Initial mass function for the stars in the region `A' of the Be~59.  Black points are from 10$^3$ realizations of the Monte Carlo simulation where each mass is moved by a random offset within its respective $\pm$ 1 $\sigma$ uncertainty. The error bars show the Poisson error in data. Black-filled circles are data from previous works \citep{pan08}.}
\label{massf}
\end{figure}

\subsection {Initial Mass Function}
The IMF represents the mass spectrum of a stellar population 
at the onset of a star formation event. The IMF of a stellar population is critical as its variation provides clues about the physical conditions of star formation processes \citep[e.g.,][]{bate09}. As young clusters are still in a very early phase of losing a significant number of members due to dynamic or stellar evolution, they are the most preferred sites to study the IMF. 

In general, the IMF is defined as the number of stars per unit logarithmic mass interval 
and is represented by a power law form : \\
$\phi$ = {{ $\rm {d}$ $N$}/{$\rm{d}$ log $M$}} $\propto$ $M^{\Gamma}$,\\
 where $M$ is the mass of a star, and $N$ is the number of stars per unit logarithmic mass interval. \citet{sal55} derived $\Gamma = -1.35$ for the stars in mass range  $0.4  < M/M_{\odot} \le 10$ in the Solar neighborhood. 

The two most widely used representations for the IMF are the segmented power-law distribution introduced by \citet{kroupa01} and the combination of log-normal (low mass) and power law (high mass) distribution by \citet{chab03} \citep[see][]{off14}. Both of these exhibit 
a broad plateau at masses from $\sim 0.1 - 1M\odot$ and then a power-law decline at higher masses.

The statistics for the member stars at different mass bins are utilized to study the IMF for the central portion of the cluster Be~59. Fig. \ref{nircmd} shows the distribution for the sources in region `A' (black dots) and control field region (blue color) in the $J$/($J$-$H$) CMD. To further quantify the level of contamination in the stellar and substellar regime, we statistically subtracted the contribution of field sources from the CMD of the cluster region using the following procedure. For a randomly selected source in the $J$, ($J$ - $H$) CMD of the field region, 
the nearest source in the cluster's $J$, ($J$ - $H$) CMD within $J$ $\pm$ 0.4 and ($J$ - $H$) $\pm$ 0.3 of 
the field source is removed \citep{pan13a,cha11}. This procedure is followed for all the sources in the field region. A statistically cleaned CMD for the cluster region was constructed using those sources in the cluster region that survived the above criteria.  
Fig \ref{nircmd} (right panel) shows the $J$ vs. ($J$ - $H$) statistically cleaned CMD of the cluster region along with the 2 Myr isochrone scaled for the adopted distance and reddening. 
The CMD clearly shows the presence of member sources in the cluster. 
This statistically cleaned CMD is later used to construct IMF for the cluster region `A'. The IMF was obtained by counting the 
number of sources in different mass bins and correcting for incompleteness by applying the completeness factor determined using the ADDSTAR routine of $DAOPHOT$-$II$ as described in Sec. \ref{compl} \citep[for details see][]{pan08,panwar18}. 
Fig. \ref{massf} shows the IMF for the cluster region `A' in the mass range 
0.03 $\le$ M/$M_\odot$ $\le$ 13. 
The data for higher masses ($>$ 1.0 M$_\odot$) have been taken from \citet{pan08}. To assess the impact of photometric uncertainties on mass determination, we incorporated random errors in the photometric measurements and utilized Monte Carlo simulations to compute the average mass and associated uncertainty of each statistically cleaned probable member. 
For the IMF estimation, we run a Monte Carlo simulation for 10$^3$ realizations where each mass is moved by a random offset within its respective $\pm$ 1 $\sigma$ uncertainty. These are plotted as black points with error bars in Fig. \ref{massf}.

As can be seen from the figure, towards the high mass range, the mass function rises up to 0.3 $M_\odot$. Nonetheless, 
we note that the low-mass IMF has been investigated for several
other SFRs, and in some cases, the results are similar to the IMF derived here. For example, a
variety of photometric and spectroscopic surveys have been completed in the highly-dense Trapezium cluster and the 
intermediate density cluster IC~348 by multiple authors \citep{hill00,mue02,mue03,sle04,luc05}. They all 
found an increasing mass function to a maximum at $0.1-0.3$ $M_\odot$ before declining 
in the BD regime. These results are very different from the mass function 
derived for the Taurus (characterized by its low gas and stellar densities), 
which peaks at 0.8 M$_\odot$ \citep{bri02,luh03a,luh04}.

\section{Discussion}
\subsection{Clues on the formation of very low-mass stellar and substellar object}
There are two main competing views on the formation of BDs. The first one assumes that the BDs form like stars, i.e., through the gravitational collapse of small, dense cores and subsequent
accretion. The alternative view is that BDs form out of the lowest mass cores \citep{reipurth01} that are ejected from unstable protostellar multiple systems and, due to the lack of enough material, end up as BDs. 

\begin{figure*}
\centering
\includegraphics[scale = 0.7, trim = 0 0 0 0, clip]{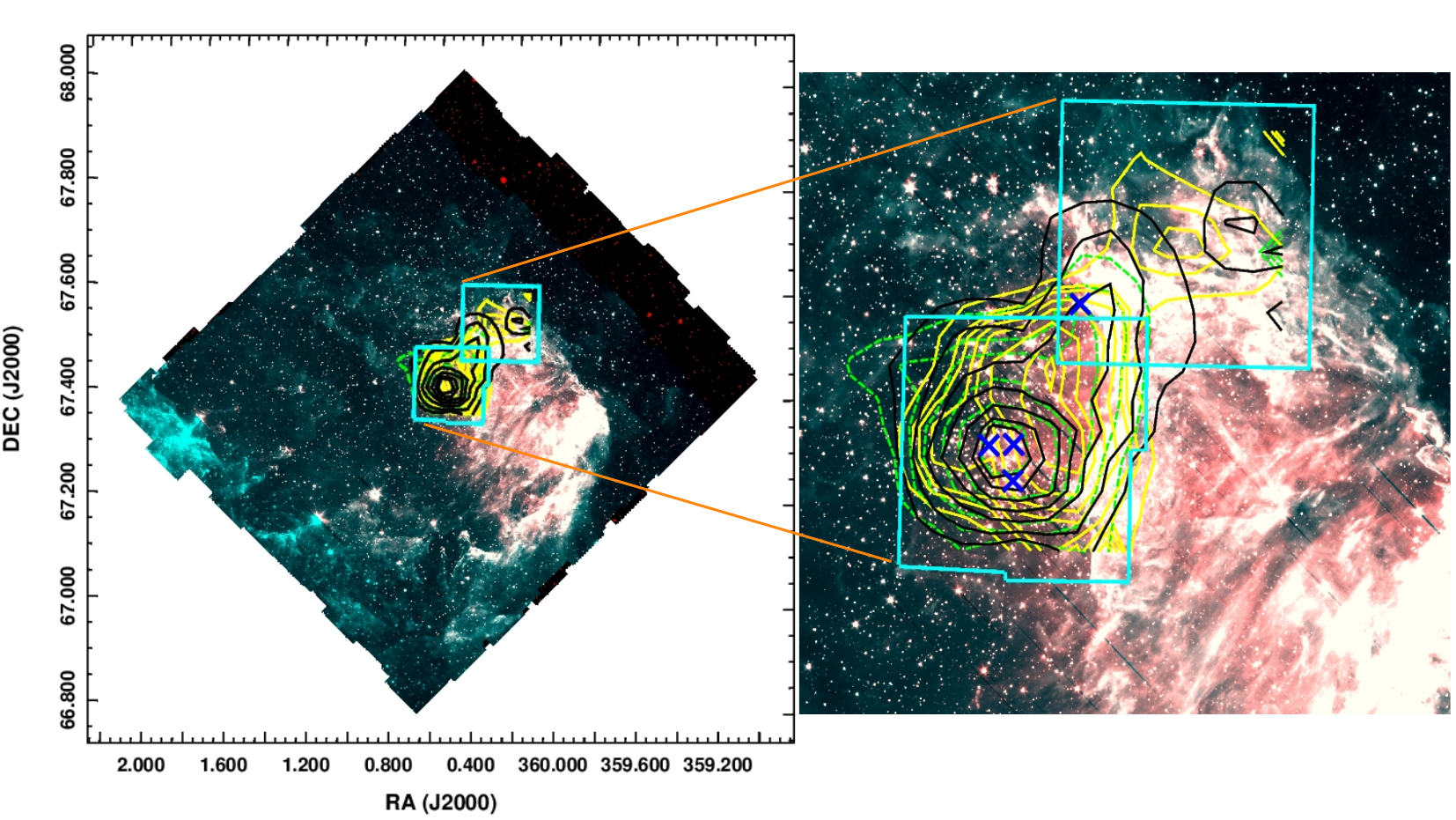}
\caption{Color-composite image (turquoise: $Spitzer$ 3.6 and  red: 4.5 $\micron$ image) of the Be 59. Regions covered with the NICS observations are shown with cyan polygons. Black contours show the density distribution of the YSO candidates in the region that were identified using various criteria discussed in Section 3. The cross symbols mark the locations of the massive O-type stars}. Dashed green and thick yellow contours show the surface density distributions of the young stellar (0.075 - 0.9 M$_\odot$) and BD ($<$ 0.075 M$_\odot$) candidates, respectively.
\label{fig9}
\end{figure*}

Investigations on the substellar-mass populations show that if substellar mass objects form as an extension of the star formation process, we shall expect an indistinguishable spatial distribution of the substellar mass population in a particular SFR \citep{parker14, parker23}.  
Recent studies show that disks are ubiquitous around the substellar objects \citep{jay03,luh12a,scholz23}. Subsequent works elaborate that the BDs' disks are capable of planet formation \citep{testi16,jung18}. The basic technique to identify circumstellar disks is excess emission from the warm dust. 
We exploit a census of very low-mass stellar and substellar sources with infrared excess and estimated the mass of these sources by dereddening each source along the reddening vector to the 2 Myr isochrone on the CMD shown in Fig. \ref{nircmd} (left panel). We examined the spatial distribution of these young sources in cluster regions `A' and `B'. For this, we have divided our young stellar and substellar candidates into two populations, one with mass in the range 1.0 - 0.075 M$_\odot$ (stellar candidates) and another in the range 0.075 - 0.04 M$_\odot$ (BD candidates). 
In Figure \ref{fig9}, we have shown surface density distributions of the young stellar and BD candidates with the dashed green and thick yellow contours, respectively. For comparison, we have also overlaid the contours for the YSO surface density distribution in the region, shown in black. The massive O-type stars are shown with the cross symbols. The surface density distribution contours for the stellar and BD populations clearly show that the density peak of the young low-mass stellar candidates nearly coincides with the cluster center. However, there is a scarcity of the BDs near the cluster center. Fig. \ref{fig9} also illustrates two density peaks of BD distribution; one is about two arcmins away from the cluster, and another is found inside the dark cloud `C2' (see Fig. \ref{fig1}), enclosed by our observed region 'B.'
Based on the spatial distribution of the stellar and BD population, a significant number of BD candidates seems unlikely to form as an extension of the star formation process.

On the other hand, the ejection scenario suggests that due to a lack of surrounding gas, the ejected sources can not accrete enough mass and end up as BDs. An important argument favoring the direct collapse scenario is the observation of some very low-mass isolated proto-BD clumps \citep{luh07}. Hence, ejected clumps, rather than finished BDs, enhance the possibility for the clumps that survive to cool and contract to form a substellar core and a disk, thus also explaining the presence of accretion disks around BDs. Furthermore, the ejection speeds of the clumps are usually $<$ 1 km $s^{-1}$, implying the ejection to a distance of about a pc in a Myr.

Therefore, with these ejection speeds, the velocity dispersion of BDs is not expected to differ much from that of YSOs, and they should remain spatially co-located within clusters forming
clumps of parsec scale for at least the typical 1–2 Myr age of YSO clusters \citep{basu12}. In the case of Be~59, the distribution of stellar and BD candidates is distinguishable. As the cluster is still very young ($\sim$ 2 Myr), the ejection scenario may not be a significant factor for the observed distribution of BD population. 

Massive members of the young stellar clusters drive stellar winds and intense ionizing radiation in the natal cloud and subsequently create H{\sc ii} regions. H{\sc ii} regions often harbor intricate structures of dusty gas, such as bright-rimmed clouds, pillars, globules, elephant trunks-like structures (ETLS) and globulettes etc., which have different sizes and shapes \citep{gah06,cha11,schn16,panwar19}. Out of these structures, the ETLSs and globulettes are of particular interest. The ETLSs are very small and elongated structures and are supposed to form very low-mass objects \citep{panwar19}. Similarly, the globulettes are small roundish objects (size $<$ 10 kAU, mass $<$ 0.1 M$_\odot$) and are considered the seeds of brown dwarfs and free-floating planetary mass objects \citep{gahm07}. These globulettes can survive up to a Myr. The H{\sc ii} region S~171, carved by the massive members of the young cluster Be~59, is a target for several triggered star formation studies \citep{ros13,get17,gahm22}. It harbors BRCs, elephant trunks, and globulettes \citep{gahm07, gahm22}. \citet{gahm22} found that in the S171 region, the high-velocity cloudlets extend over a larger radius and are less massive than the low-velocity cloudlets. Therefore, some of the substellar candidates in the Be~59 may result from the in-situ formation in the globulettes or ETLSs, which are now either dissolved or not detectable in the molecular line observations. The observed spatial distribution of the BD candidates in the regions 'A' and 'B' in Fig. \ref{fig9} also indicates an anisotropic distribution, i.e., most of the BDs are distributed toward the cloud `C2' in region `B', which is a potential triggered star formation site due to radiative feedback from the massive star in the region. We note that the current observations do not cover the whole cluster region. Still, the present analysis indicates that the radiation feedback from the massive stars in the region may have some influence on the formation of BD candidates. However, deep NIR observations of the wider area will be helpful in shedding more light on this subject.

\begin{figure*}
\centering
\includegraphics[scale = 0.46, trim = 0 0 0 0, clip]{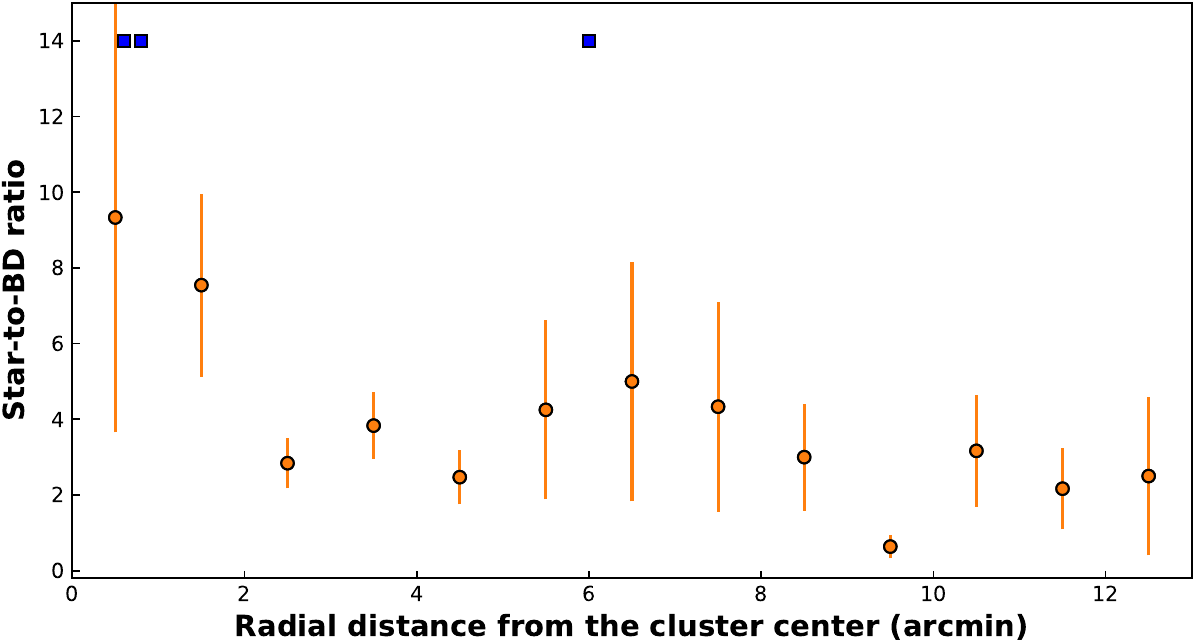}
\includegraphics[scale = 0.45, trim = 0 0 0 0, clip]{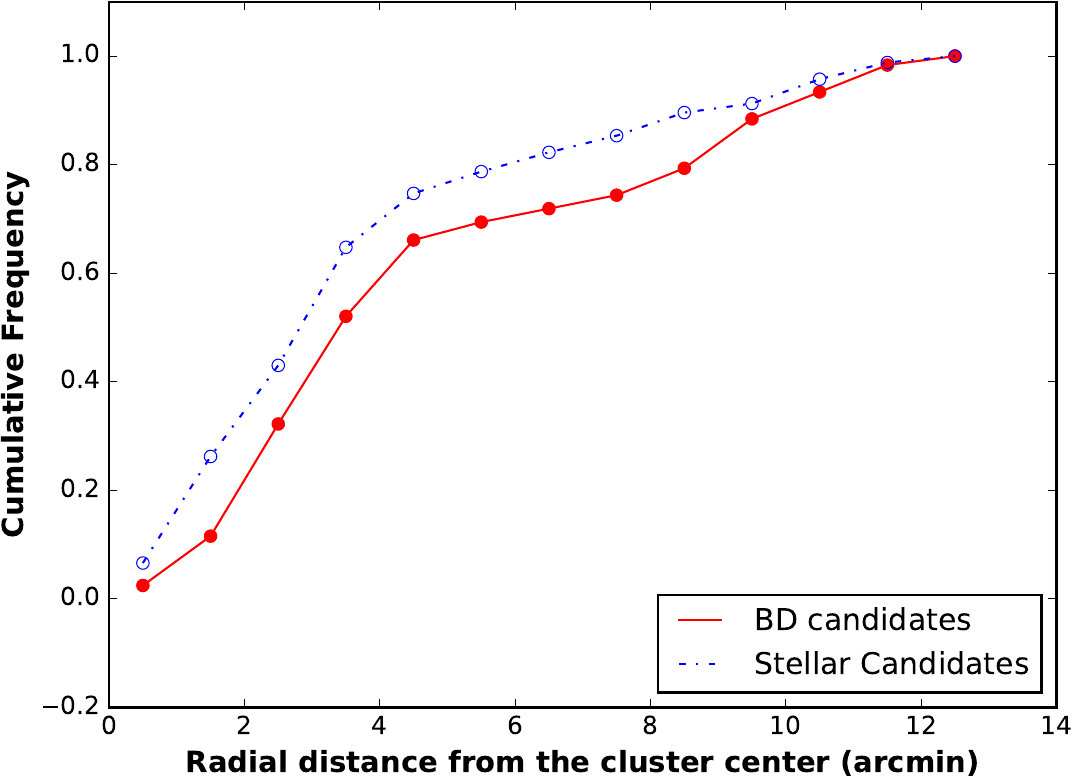}
\caption{Left panel: The variation of star-to-BD ratio with the radial distance from the cluster center. The blue-filled squares represent the massive `O' type stars. Right panel: The cumulative distribution of the BD and stellar candidates as a function of the radial distance from the cluster center. }
\label{fig10}
\end{figure*}
\subsection{Environmental influence on the BD formation}
The environmental conditions in a star-forming region (SFR) can impact BD formation. For example, numerical simulations suggest that BDs are preferentially formed in regions of high stellar densities through the gravitational fragmentation of infalling gas into stellar clusters \citep{bonnell08} or, by ejection of embryos from the reservoir of natal material \citep{bonnell08,bate12}. Furthermore, the turbulent fragmentation framework also predicts that an enhancement in density should lead to an increase in the number of BDs \citep{padoan02,lormax16}. 
The other important factor influencing BD frequency is the presence of massive stars in the cluster. During their initial stages, low-mass objects accrete matter from their outer envelope. The powerful ionization fronts of the massive OB star(s) can erode the outer layers of a pre-stellar core, leaving a small fragment that can only form a substellar object \citep{whitworth04,fang16}. 
\subsubsection{IMF in the substellar mass regime}
To examine the influence of the environment in the production of the substellar mass, we have compared the IMF of the cluster Be~59 with other nearby young clusters in the stellar and substellar mass regime. 
Though a single slope for the IMF in the mass range 0.2 $\le$ M/$M_\odot$ $\le$ 13 can be fitted with $\Gamma$ = -1.35 (Salpeter Value), towards the low-mass regime ($<$ 0.2 M$_\odot$), the resultant $\Gamma$ value becomes shallower. For comparison, we have also plotted segmented power 
law from \citet{kro01}, normalized at 0.5 M$_\odot$, with a slope $\Gamma$ = -1.35 for mass $>$ 0.5 M$_\odot$; $\Gamma$ = -0.3 for the mass range 0.5 - 0.08 M$_\odot$ and $\Gamma$ = 0.7 for mass $<$ 0.08 M$_\odot$. A log-normal IMF from \citet{chab03} is also shown as a dashed curve in Fig. \ref{massf}. 
The IMF slopes $\Gamma$ = -1.30 $\pm$ 0.07 and $\Gamma$ = -1.31 $\pm$ 0.11 in the mass range 0.2 - 13 M$_\odot$ and 0.4 - 13 M$\_odot$, respectively, are in agreement with the Salpeter slope. For the masses above 0.2 M$_\odot$, the log-normal part of the Chabrier IMF does not seem to represent the
IMF very well. For the masses below 0.2 M$_\odot$, a single power law slope seems to describe IMF  
to both the low-mass stellar and substellar regimes. We found that in a mass range 0.04 - 0.4 M$_\odot$, the IMF slope, $\Gamma$ is $\sim$ 0.01 $\pm$ 0.18. In some of the recent works on the substellar IMF estimation, the IMF is described as the number of stars within a given mass interval, with $\alpha$ defining the slope for the mass range of interest, where $dN/ dM$ $\propto$ $M^{-\alpha}$. We used the relation  $\alpha$ = 1 - $\Gamma$ to compare our IMF results with the other nearby clusters. The value of the IMF in the mass range 0.04-0.4 M$\odot$ is comparable to that obtained for the cluster NGC 2244 but slightly steeper than the one in RCW 38 ($\alpha$ $\sim$ 0.7) in the similar mass range \citep{muzic19}. 
However, the substellar IMF slope of the Be~59 lies within the range of IMF slopes for the nearby clusters ($\alpha$ $\sim$ 0.5 - 1.0).

The region `B' consists of the cloud `C2' and may be a probable triggered candidate. For comparison, we also estimated the IMF slopes for the stars in region `B' by adopting a similar approach as in the case of the region `A'. The IMF slope for the region `B' is $\Gamma$ = - 1.05 $\pm$ 0.14 in the mass range 0.4 - 13 M$_\odot$. In contrast, it is found to be $\Gamma$= -0.29 $\pm$ 0.12 in the mass range 0.04 - 0.4 M$_\odot$, suggesting that there are relatively more BDs in region `B' as compared to the region `A'.  
\subsubsection{Star-to-BD ratio}
The star-to-brown dwarf number ratio is frequently utilized as an indicator to evaluate whether the number distribution of very-low-mass stars and BDs is similar across different regions or varies based on the environment. 
In most of the observational works related to the star-to-BD ratio estimation, the stellar mass range is adopted as $\sim$ 0.075 - 1 M$_\odot$ whereas the mass range for the BDs is taken as 0.075 M$_\odot$ to 0.03 M$_\odot$. In the present work, we have adopted similar mass ranges to estimate the star-to-BD number ratio in the central portion of the cluster (region `A') and near the region `B' that come out to be $\sim$ 3.6 and 2.2, respectively. Most recent works show that the value of the star-to-BD number ratio ranges from 2 - 5. For example, \citet{muzic17} found a value of $\sim$ 3 for the RCW 38 and \citet{scholz13} found $\sim$ 1.9 - 2.4 for NGC~1333 and 2.9 - 4 for IC~348. In the case of Chamaeleon-I and Lupus-3, the ratio was reported as $\sim$ 2.5 - 6 \citep{muzic15}. \citet{muzic19} found that even though the stellar densities of the RCW 38, ONC, and NGC 2244 are pretty different, the BD frequencies of these regions are somewhat similar, indicating that stellar densities might not have a significant role on the BD formation. In the case of Be~59, the stellar density in the central portion of the cluster is $\sim$ 150 pc$^{-2}$, which is lower than the RCW~38 and the ONC but similar to NGC 2244 and NGC~1333, emphasizing the insignificant influence of the stellar density on BD frequency.
\subsubsection{Radial variation of star-to-brown dwarf ratio}
We examined the variation of the star-to-BD ratio as a function of radial distance from the cluster center in the observed region, shown in Fig. \ref{fig10} (left panel). The respective error bars in the ratio are also shown. The massive O-type stars in the cluster region (see Sec. 1) are shown as filled squares. Although the figure manifests a higher star-to-BD ratio near the cluster's center, we do not see any significant variation within uncertainties. Recently, \citet{almendros23} analyzed the substellar content of the cluster NGC~2244 using spectroscopically confirmed members and found that BDs were closer to massive stars than low-mass stars. However, in the case of Be~59, within uncertainties, we do not see significant variation in the star-to-BD ratio near the massive stars. A dip in the ratio at $\sim$ 9 arcmin radial distance is around the dark cloud, indicating the higher number of substellar sources possibly formed due to the radiation feedback from the massive stars. 

We also examined the mass segregation in the low-mass stellar and substellar mass regimes. Fig. \ref{fig10} (right panel) shows the cumulative distribution of the BD and stellar candidates as a function of the radial distance from the cluster center. Fig. \ref{fig10} (right panel) clearly shows that the BD candidates are generally away from the cluster center compared to stellar candidates, which suggests mass segregation in the low-mass stellar and substellar regimes. We performed the two-sample Kolmogorov-Smirnov (KS) test to investigate the null hypothesis that the two populations were drawn from the same distribution. The KS test yields a p-value of 0.01, which indicates that the two populations significantly differ. This result is similar to that found by \citet{panwar18} for the low-mass stars (0.3 - 1.5 M$_\odot$). As the cluster is still young (age $\sim$ 2 Myr), the observed mass segregation in the cluster may be primordial. 

\section{Summary and Conclusions}
We present the analyses of Gaia DR3 and deep $JHK$ observations of the central portion of the cluster Berkeley 59 using the 
3.58-m TNG, which are the deepest NIR ($J$ $\sim$ 20 mag) observations so far for the region. The main results of the present work are as follows: 
\begin{enumerate}
    \item The VPD constructed using the Gaia DR3 data for the stars within cluster extent ($\sim$ 10 arcmin) remarkably differentiates the cluster members from the field stars. The mean proper motion of the cluster is $\mu$$_\alpha$cos($\delta$) $\sim$ -0.63 mas yr$^{-1}$, $\mu$$_\delta$ $\sim$ -1.83 mas  yr$^{-1}$. The mean distance of the cluster obtained using the distances of the cluster members estimated from the parallax values given in Gaia data \citep{bailler21} agrees with the photometric distance of $\sim$ 1 kpc. 
    \item We constructed NIR CMD for the sources in the region `A' and `B'. Adopting the visual extinction, A$_V$ = 4 mag, age of 2 Myr and distance of $\sim$ 1 kpc towards the cluster region, the CMD for the cluster members show that our NIR data are the deepest available data for the cluster and reach below the hydrogen burning limit. The distribution of the sources in the CMD shows a clear distinction between the cluster and field population toward the brighter end ($J$ $<$ 16 mag). In contrast, it is challenging to distinguish cluster members from field stars towards the fainter end.
    \item  The IMF of the central portion of the cluster obtained using statistical subtraction of the field star contribution from the cluster field, in the mass range of 0.04 - 0.4 M$_\odot$, is found to be $\Gamma$ $\sim$ 0.01, similar to that found for other nearby young SFRs.  
    \item The spatial distribution of the BD and stellar candidates of the cluster suggest the significance of radiation feedback from massive stars in the BD formation in the case of Be~59. 
    \item The KS test for the distribution of stellar and BD candidates shows that these two populations significantly differ, and stellar candidates are closer to the cluster center compared to the BDs, indicating mass segregation in the cluster toward the substellar mass regime. 
    \end{enumerate}

\section*{Acknowledgements}
We are thankful to the anonymous referee for the valuable suggestions that improved the scientific content of the present work. NP is grateful to the late Dr. A. K. Pandey for all the support and encouragement during the initial stage of the project. NP thanks Prof. P. Battinelli and Prof. F. Mannucci for their invaluable discussion and expertise on the SNAP pipeline. NP and RC acknowledge the financial support received through the SERB CRG/2021/005876 grant. DKO acknowledges the support of the Department of Atomic Energy, Government of India, under Project Identification No. RTI 4002. This work is based on observations made with the TNG operated on the island of La Palma by the Fundación Galileo Galilei of the INAF (Istituto Nazionale di Astrofisica) at the Spanish Observatorio del Roque de los Muchachos of the Instituto de Astrofisica de Canarias. This publication makes use of data from the Two Micron All Sky Survey (a joint project of the University of Massachusetts and the Infrared Processing 
and Analysis Center/ California Institute of Technology, funded by the 
National Aeronautics and Space Administration and the National Science 
Foundation), archival data obtained with the {\it Spitzer Space Telescope}
(operated by the Jet Propulsion Laboratory, California Institute 
of Technology, under contract with NASA. This work has made use of data from the European Space Agency mission {\it Gaia} (https://www.cosmos.esa.int/gaia), processed by the {\it Gaia} Data Processing and Analysis Consortium (DPAC). Funding for the DPAC
has been provided by national institutions, in particular, the institutions participating in the {\it Gaia} Multilateral Agreement. 

\vspace{5mm}





\bibliography{r}{}
\bibliographystyle{aasjournal}

\end{document}